\documentclass[twocolumn,amssymb,nobibnotes, aps, prl]{revtex4-1}
\usepackage{amsmath,bm}
\usepackage{graphicx}
\usepackage{hyperref,epstopdf}

\begin{document}

\title{Realization of Hofstadter's  butterfly and a one-way edge mode in a polaritonic system}%

\author{R. Banerjee$^1$}\email[Corresponding author:~]{RIMI001@e.ntu.edu.sg}
\author{T.C.H. Liew$^1$}\email[Corresponding author:~]{tchliew@gmail.com}
\author{O. Kyriienko$^2$}

\affiliation{$^1$Division of Physics and Applied Physics, School of Physical and Mathematical Sciences, Nanyang Technological University, Singapore 637371, Singapore}

\affiliation{$^2$Nordita, KTH Royal Institute of Technology and Stockholm University, Roslagstullsbacken 23, SE-106 91 Stockholm, Sweden}


\begin{abstract}
We present a scheme to generate an artificial gauge field for the system of neutral bosons, represented by polaritons in micropillars arranged into a square lattice. The splitting between the two polarizations of the micropillars breaks the time-reversal symmetry (TRS) and results in the effective phase dependent hopping between cavities. This can allow for engineering a non-zero flux on the plaquette, corresponding to an artificial magnetic field. Changing the phase, we observe a characteristic Hofstadter's  butterfly pattern and the appearance of chiral edge states for a finite size structure. For long-lived polaritons, we show that the propagation of wave packets at the edge is robust against disorder. Moreover, given the inherent driven-dissipative nature of polariton lattices, we find that the system can exhibit topological lasing, recently discovered for active ring cavity arrays. The results point to a static way to realize artificial magnetic field in neutral spinful systems, avoiding the periodic modulation of the parameters or strong spin-orbit interaction. Ultimately, the described system can allow for high-power topological single mode lasing which is robust to imperfections.
\end{abstract}

\maketitle
\section{Introduction}
In a system of charged particles the application of a magnetic field qualitatively changes its behavior, and usually is responsible for intriguing physical effects. For instance, consider electrons in a periodic potential, which in the absence of an electromagnetic field gives rise to Bloch bands. In the presence of perpendicular uniform magnetic field the spectrum splits into highly degenerate Landau levels. It was predicted that the interplay between a periodic potential and magnetic field for the two dimensional electron gas leads to the emergence of a self-similar fractal energy spectrum, known as Hofstadter's  butterfly~\cite{Hofstadter}. It corresponds to the plot of the allowed and forbidden energies as a function of magnetic flux. For a rational value of the normalized magnetic flux $\beta=p/q$ (where $p$ and $q$ are co-prime integers) each Bloch band splits into $q$ subbands. These energy bands deform into Bloch bands only after closing the fractal gaps between them, admitting a topological characterization in the system~\cite{ThoulessKohmoto}. This also leads to the appearance of topologically protected edge states. Ultimately, the application of magnetic field to the strongly correlated system may give rise to topological order, leading to the fractional quantum Hall effect~\cite{Tsui,Laughlin}.
 
While the described physics naturally emerges for electrons and other charged particles, systems of neutral particles do not exhibit the same behavior, limiting their possible scope of application. This posed the question of the possibility to generate an \textit{artificial} gauge field~\cite{Dalibard}, where the effect of the field is simulated by some means. For continuous systems, examples include the generation of magnetic field for cold atoms gases using rotation~\cite{Fetter} and laser illumination~\cite{Dalibard,Spielman}. For lattice systems, different microwave~\cite{Raghu2008}, optical~\cite{Hafezi2011,Umucalilar2011,Fang2012,MittalFan,SchmidtKessler}, and cold atom~\cite{ChinMueller,Umucalilar,Yilmaz} setups were considered recently, and recipes for artificial gauge field generations were proposed. Up to date, Hofstadter's  butterfly was demonstrated using engineered superlattice structures with microwave resonators~\cite{Wang2009}, bilayer graphene~\cite{DeanWang,Ponomarenko}, optical ring microresonators~\cite{Hafezi2013}, and cold atom lattices~\cite{ChinMueller,Miyake, Aidelsburger,Stuhl2015}. Finally, the Harper Hamiltonian was simulated using a chain of superconducting circuits \cite{Roushan2017}, promising the way towards implementation of magnetism in the interacting systems.

The described systems and approaches to observe fractal Hofstadter butterfly behavior can be classified into several categories. First, the lattice models with phase dependent hopping can be divided into systems where TRS is broken (integer quantum Hall effect type physics) and not broken (spin Hall effect type physics). In the former case, there is a unique chiral edge state associated to the boundary, while in the latter case two co-propagating channels exist, albeit for different spin components. From the point of realization, the butterfly behavior may be posed by magnetooptical effects, time-dependent modulation of the tight-binding coupling \cite{Fang2012}, and phase-dependent hopping on a spinful lattice~\cite{Hafezi2011}. We note that while for microwave and cold atom systems the first two approaches have allowed to break TRS explicitly, for optical systems (e.g. silicon resonators) this was shown to be notoriously difficult. Thus, so far only spin Hall physics with non-broken TRS was studied~\cite{Hafezi2013}.

One of the promising platforms where the realization of artificial gauge fields and the associated Hofstadter's butterfly is missing is the system of  exciton-polaritons. Polaritons, quasiparticles which arise due to the strong coupling between photons and quantum well excitons, have attracted growing interest in the last few decades \cite{CarusottoCiuti, DengHaug}. Due to their photonic part, polaritons have very small effective mass and long coherence length which enable them to show Bose-Einstein condensation~\cite{Kasprzak,Byrnes} and superfluidity~\cite{AmoLefrere,Sanvitto} even at room temperature~\cite{Lerario}. Polaritons exhibit strong nonlinearity due to their excitonic part, which makes them a potential candidate for  optoelectronic applications such as logic gates~\cite{Anton, Kyriienko, AntonLiew}, optical circuits~\cite{Ortega}, optical transistors~\cite{Ballarini} etc. \par

In traditional condensed matter systems, as an atomic spacing is very small (of the order of an angstrom), one needs a large magnetic field (thousands of Tesla to get one flux quantum per unit cell) to observe Hofstadter's  butterfly~\cite{Albrecht2001}. Alternatively, the self-similar pattern can be observed in semiconductor nanostructures hosting high-mobility two-dimensional electron gas, with the presence of superlattice~\cite{PfannkucheGerhardts,GerhardtsPfannkuche1996}; in the polaritonic system the periodicity is on the scale of microns, and from the experimental perspective the system favours the possibility to observe Hofstadter's  butterfly. However, due to the charge neutrality of the polaritons, it is not easy to affect the orbital motion significantly and to break TRS directly~\cite{Koch} by applying magnetic field. Thus, in order to mimic the effect, one needs to synthesize an artificial magnetic field in the system. For  continuous polaritonic systems the artificial magnetic field was implemented using the magnetoelectric Stark effect~\cite{Lim_Imamoglu}. 

In this paper we consider theoretically a system of coupled micropillars with polarization splitting at each site, along with spin-dependent coupling. This breaks time-reversal symmetry even in the absence of a real magnetic field. The splitting is realized in a way that effectively couples different lattice sites $i$ and $i'$ with a phase factor $\phi_{i,i'}$. Controlling $\phi_{i,i'}$ on the square lattice, we show that the analog of Hofstadter's  butterfly for the spectrum of the system and chiral edge modes can be observed. The latter is demonstrated to avoid backscattering from the disorder, potentially leading to  unidirectional transport in the polaritonic system. 

Finally, considering the driven-dissipative nature of exciton-polaritonic lattices, we show that the combination of the topological transport and gain in the system can lead to the single mode lasing associated to an edge mode. This corresponds to the recently discovered topological insulator lasing, which was described theoretically in Ref. [\!\!\citenum{Harari_Chong_18_theory}]  and experimentally confirmed in Ref. [\!\!\citenum{Bandres_18_Exp}] for ring cavity arrays. We demonstrate that polariton lattices can host a lasing mode corresponding to a chiral edge state, which has a positive gain and is unhampered by disorder. This opens the way towards polaritonic lasers of largely increased efficiency and large gain.

\begin{figure}[t]
\centering
\includegraphics[width=\linewidth]{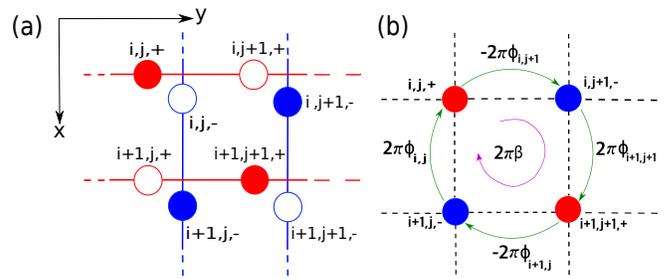}
\caption{(a) Full model: Schematic diagram of a square lattice formed by coupled exciton-polariton elliptical micropillars that support two kinds of polarizations ($\sigma=\pm 1$) with the splitting $\Delta$. Here, filled circles denote relevant (\textit{main}) modes, and empty circles denote auxiliary modes. The angle of the micropillar polarization axis at site ($i,j$) is $\phi_{i,j}$. The tunneling (overlap) between nearest micropillars is allowed only for polaritons of the same polarization. The two spin polarizations of a micropillar interact differently, one hopping horizontally and the other one vertically, while other channels are closed by the polarized potentials. The red lines describe the direct tunneling between neighbors of polarization $\sigma=+1$, while blue lines denote the direct tunneling between neighboring pillars of polarization $\sigma=-1$, both described by the hopping amplitude $J$. (b) Reduced model, where only main logical modes are shown. The dashed lines represent the complex hopping associated with magnetic vector potential $\mathbf{A}$ (see the discussion in the text), which is responsible for the realization of an artificial magnetic field in the system. Polaritons can hop in four orthogonal directions, and pick up non-zero phases which are direction dependent. The phase per plaquette is uniform and equal to $2\pi\beta$.}
\label{fig:combine_model}
\end{figure}

\section{The model}
We consider an array of coupled exciton-polariton elliptical micropillars which form a square lattice, as shown in Fig.~\ref{fig:combine_model}(a). In this vein, several experiments up to date were performed for the coupled micropillars or arrays of micropillars~\cite{Rodriguez,Milicevic,Sala,Winkler,Klembt,ZhangBrodbeck2015}.  Each micopillar has two kinds of polarization ($\sigma =\pm 1$).  We define the lattice in terms of two sets of modes corresponding to alternating spin polarizations in successive micropillars. The first set of modes (represented by filled circles in Fig.~\ref{fig:combine_model}a) will be called the (main)  \textit{logical} modes; the second set of modes (represented by empty circles in Fig.~\ref{fig:combine_model}a) will be called the \textit{auxiliary} modes. The hopping between only the same polarization of two neighboring pillars is allowed. We consider modes with $\sigma=+1$ tunnelling horizontally, whereas modes with $\sigma=-1$ tunnel vertically. We note that this spin dependent hopping corresponds to a breaking of time reversal symmetry, which will later be shown to give rise to an artificial gauge field. This kind of potential landscape can be created by non-resonant excitation with a localized optical pump which induces a spin-dependent potential that can selectively gate different spin polarizations \cite{Gao}. Recent work also shows that alternating patterns of spin polarization may form spontaneously under near resonant excitation, offering a further alternative for engineering spin-dependent coupling for our lattice \cite{Gavrilov}. In the tight-binding approximation with nearest-neighbor coupling the Hamiltonian can be written as 

\begin{align}
\hat{H} =& \sum_{i,j,\sigma}\Big[-\Delta e^{-i2\pi\phi_{i,j}\sigma} \hat{a}^{\dagger}_{i,j,\sigma }\hat{a}_{i,j,\bar{\sigma} }\nonumber\\
&+J \hat{a}^{\dagger}_{i,j,\sigma }\hat{a}_{(i-\delta_{1,\bar{\sigma}}),(j-\delta_{1,\sigma}),\sigma }\nonumber\\
&+J \hat{a}^{\dagger}_{i,j,\sigma }\hat{a}_{(i+\delta_{1,\bar{\sigma}}),(j+\delta_{1,\sigma}),\sigma }+\text{h.c}\Big],
\label{hamiltonian}
\end{align}
where the operators $\hat{a}^{\dagger}_{i,j,\sigma} ~(\hat{a}_{i,j,\sigma} )$ create (annihilate) an exciton-polariton at site ($i,j$) with circular polarization $\sigma$. The first term in Eq.~(\ref{hamiltonian}) describes the coupling between the cross-polarized polaritons in the same micropillars, leading to the splitting between $\sigma = \pm 1$ components. It has been shown previously in polariton microwires that their asymmetry in the horizontal and vertical directions results in a polarization splitting~\cite{Dasbach}. Similarly, the micropillars could be grown with chosen asymmetry, which we model by  the linear polarization splitting $\Delta$ and an angle $\phi$. The next two terms describe the tunneling between nearest neighbors having the same polarization as drawn by solid lines in Fig.~\ref{fig:combine_model}(a). Using the described model, we can write down the equation of motion for the polariton field at site ($i,j$) as 

\begin{align}\label{1st_order_eq}
{i\hbar}\frac{\partial\psi_{i,j,\sigma}}{\partial t} =& \Big(-\Delta e^{-i2\pi\phi_{i,j}\sigma} \psi_{i,j,\bar{\sigma}}+J \psi_{(i-\delta_{1,\bar{\sigma}}),(j-\delta_{1,\sigma}),\sigma}\nonumber\\
&+J \psi_{(i+\delta_{1,\bar{\sigma}}),(j+\delta_{1,\sigma}),\sigma}\Big),
\end{align}
where we observe the onsite coupling through polarization splitting (first term on r.h.s.), as well as intersite tunneling terms. Our goal is to generate an effective phase-dependent interpillar tunneling between main logical modes, which is mediated by the polarization splitting. For this, the auxiliary modes, chosen to be \{$\psi_{i,j,-}$; $\psi_{i+1,j+1,-}$; $\psi_{i,j+1,+}$; $\psi_{i+1,j,+}$\}, have to be eliminated. 

The elimination can be done by taking the second derivative at the l.h.s. of Eq. (\ref{1st_order_eq}) and substituting the evolution of the neighbors. Then, the evolution of site ($i,j$) can be described by a quadratic equation,

\begin{align}\label{2nd_order_eq}
\frac{\partial^{2}\psi_{i,j,\sigma}}{\partial t^{2}}=& \Big(-\frac{\Delta^{2}+2J^{2}}{\hbar^{2}} \psi_{i,j,\sigma} \nonumber\\
&+\frac{J\Delta e^{-i2\pi\sigma\phi_{i,(j-\delta_{1,\sigma})}}}{\hbar^{2}}\psi_{i,j-1,\bar{\sigma}}\nonumber\\
&+\frac{J\Delta e^{-i2\pi\sigma\phi_{(i+\delta_{1,\bar{\sigma}}),j}}}{\hbar^{2}}\psi_{(i+1),j,\bar{\sigma}}\nonumber\\
&+\frac{J\Delta e^{-i2\pi\sigma\phi_{i,(j+\delta_{1,\sigma})}}}{\hbar^{2}}\psi_{i,j+1,\bar{\sigma}}\nonumber\\
&+\frac{J\Delta e^{-i2\pi\sigma\phi_{(i-\delta_{1,\bar{\sigma}}),j}}}{\hbar^{2}}\psi_{i-1,j,\bar{\sigma}}\nonumber\\
&-\frac{J^{2}}{\hbar^{2}}\psi_{(i-2\delta_{1,\bar{\sigma}}),(j-2\delta_{1,\sigma}),\sigma}\nonumber\\
&-\frac{J^{2}}{\hbar^{2}}\psi_{(i+2\delta_{1,\bar{\sigma}}),(j+2\delta_{1,\sigma}),\sigma}\Big) \nonumber ,\\
\end{align}
and we note that no approximation was made so far. Eq.~(\ref{2nd_order_eq}) demonstrates that the system decouples into two separate parts, corresponding to the main and auxiliary modes [filled and empty circles shown in Fig.~\ref{fig:combine_model}(a)]. Considering the filled circles, the model is reduced to Fig.~\ref{fig:combine_model}(b). The first term in Eq.~(\ref{2nd_order_eq}) describes the on-site energy. The middle four terms describe the tunneling with four orthogonal sites in the reduced model, which are phase dependent. These are the plaquette terms which are required to generate an artificial magnetic field. Finally, the last two terms denote the interaction with the next nearest neighbors (NNN) in the reduced model, which are inherent to our effective model, and are typically absent in the other geometries (e.g. cold atomic lattices with modulated hopping \cite{Aidelsburger}). In the following we highlight the differences, and show the limit where NNN coupling can effectively be neglected.

As a result of polariton hopping from one site to another, the spatially-dependent complex tunneling leads to the accumulation of a non-trivial phase. This tunneling is directionally dependent. It means that when particles hop between two sites clockwise, the value of the accumulated phase will be the negative of the value when it hops anticlockwise. Associating a gauge potential $\mathbf{A}$ to the phase $\phi_{c}=(q/h)\int_{c}\mathbf{A} \cdot d\mathbf{r}$, this introduces an effective artificial magnetic field in the system which is given by $\text{B}=(1/a^2)\oint_{\text{plaquette}}\mathbf{A} \cdot d\mathbf{r}$, where $a^{2}$ is the area of the plaquette. Physically, the phases $\phi_{i,j,\sigma}$ are dependent on the chosen polarization axis of the micropillars and we choose the phases in such a way that polaritons accumulate a uniform phase ($2\pi\beta$) after completing each plaquette. Therefore, the magnetic flux per lattice unit cell in units of the magnetic flux quantum is set to be $\beta$. As a result we can realize a uniform artificial magnetic field in the system.
\section{Results}
\subsection*{Hofstadter butterfly}
\begin{figure}[htb]
\centering
\includegraphics[width=\linewidth]{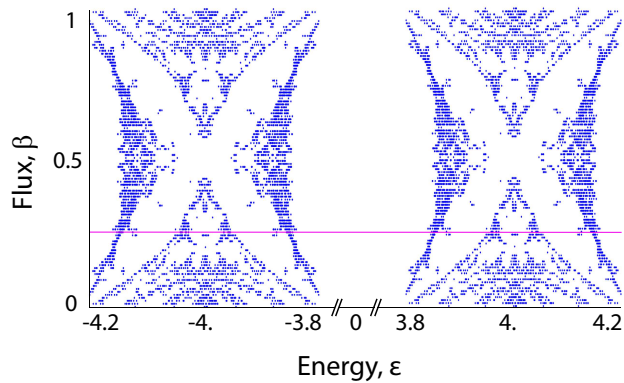}
\caption{Hofstadter butterfly spectrum. Energy $\varepsilon$ (stated in meV) is shown as a function of dimensionless parameter $\beta $, which is the effective magnetic flux per lattice unit cell in terms of magnetic flux quantum. We consider a $10\times10$ lattice with periodic boundary condition to study the bulk properties, and effectively subject the system to a uniform magnetic field by making the flux per plaquette uniform ($2\pi \beta$). Here, we consider $\Delta=4$ meV and $J=0.1$ meV. For each $\beta$ we calculate the density of states (DOS), find the maximum value (say $M$), and plot the points corresponding to DOS being greater than $0.36 \times M$. The choice of $\beta=1/4$ is shown by the pink line.}
\label{fig:butterfly}
\end{figure}

First, we consider the full model [see Fig.~\ref{fig:combine_model}(a)] with a square lattice of $(10\times 10)$ micropillars, corresponding to $200$ polarization modes. Imposing the toric boundary conditions we mimic the effect of an infinite lattice. Solving the eigenvalue equation for the system, the energy spectrum ($\varepsilon$) of the single particle as a function of normalized magnetic flux $\beta$ is given by  Fig.~\ref{fig:butterfly}. Notably, due to the structure of dynamical equations, instead of solving the full model, by diagonalizing the reduced model under periodic boundary condition, we can also obtain the same spectrum of two identical Hofstadter butterfly patterns which are symmetric about the $\varepsilon=0$ axis. The tunneling amplitude $J$ controls the width of the spectrum ($\simeq 4J$), whereas the limits of the eigenvalues corresponding to each magnetic flux are symmetric around $\varepsilon=\Delta$.  At the same time, compared to the commonly described Hofstadter butterfly, the wings of each are skewed. The symmetry of the spectrum depends on the ratio of $J$ to $\Delta$, and close-to-symmetric butterflies can be observed in the limit $J/\Delta \rightarrow 0$. So although the NNN terms are present in the equations, by taking the ratio small ($J/\Delta=0.025$) the effect is negligible and we observe the self-similar spectrum. From onwards, we focus on the region with $\beta=1/4$ only.

\subsection*{Band structure}
\begin{figure}[htb]
\centering
\includegraphics[width=\linewidth]{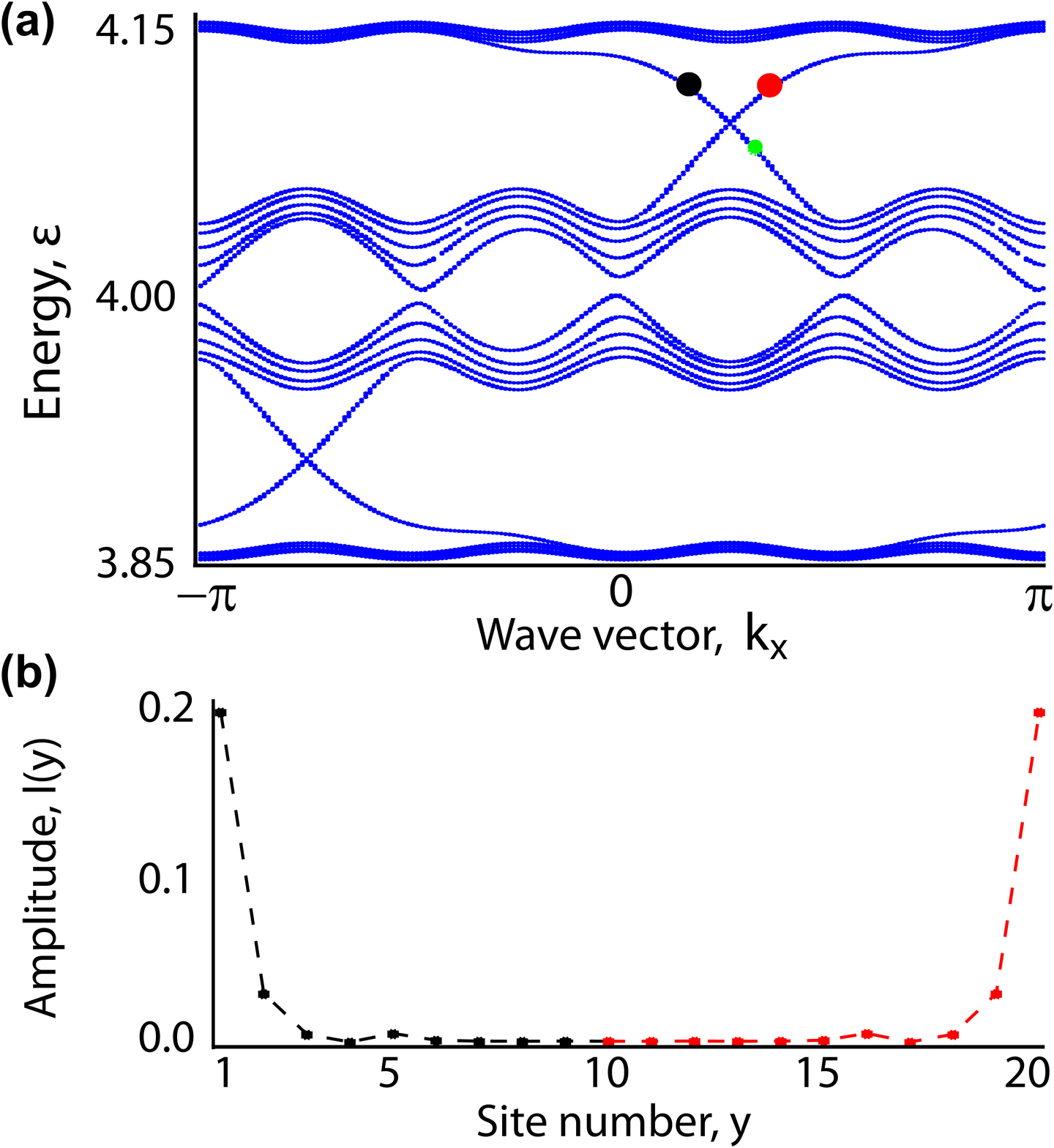}
\caption{(a) Energy diagram of the finite rectangular lattice ($200\times 20$) formed by micropillars. As we consider $\beta=1/4$, we have four bands and associated edge states. The upper edge state corresponds to the propagation clockwise along the edges, whereas the lower edge state corresponds to the propagation anticlockwise along the edges. The energy and wavevector corresponding to the green dot is used in section C.  (b) Field intensity of the two edge modes indicated by the black and red dots in Fig.~\ref{fig:band_structure}(a). The fields are localized on the two edges and decay exponentially into the bulk.}
\label{fig:band_structure}
\end{figure}

In the previous section, accounting for the effect of infinite lattice by taking the toric boundary condition, we obtain only the bulk states which correspond to the energy bands separated by the band gaps. However, for the finite lattice, the breaking of translational symmetry leads to the appearance of distinct edge states. In the case of the lattice system subjected to the artificial magnetic field, the bulk states are connected by edge modes, which are propagating unidirectionally (clockwise or anticlockwise) along the edges. Here we take a finite rectangular lattice of dimension $N_{x}\times N_{y}$ [see Fig.~\ref{fig:combine_model}(a)] corresponding to the $2 \times N_{x}\times N_{y}$ polarization modes and after diagonalizing we consider the modes with positive polarizations. Taking two dimensional discrete Fourier transforms of the corresponding eigenstates, and summing over $k_{y}$, we obtain the energy spectrum as a function of $k_{x}$ [see Fig.~\ref{fig:band_structure}(a)]. We consider the case $\beta=1/4$, so the acquired phase per plaquette is $2\pi\beta=\pi/2$ implying that the magnetic unit cell consists of four lattice unit cells and as a result the single particle Bloch band splits into four subbands. Fig.~\ref{fig:band_structure}(b) shows the integrated intensity of the states marked by black and red spots in Fig.~\ref{fig:band_structure}(a) in the $y$ direction. Their spatial structure confirms the edge-type behavior for the states.

\subsection*{Edge state transport}
Next, we simulate the transport properties of the system. To illustrate the propagation of the edge mode, we excite one of the edges by introducing a coherent pulse of the form 

\begin{align}\label{pulse}
F =& F_{0}\exp[-((x-x_{0})^2+(y-y_{0})^2)/L^{2}] \nonumber\\
  & \times \exp[-(t-t_{0})^2/\tau^{2}] \exp[i(k_{p}x-\omega_{p}t)],
\end{align}
where $F_{0}$ is the amplitude of a Gaussian pulse of width $L$, which is centered on the edge at ($x_{0}$,$y_{0}$). We choose $k_{p}$ and $\omega_{p}$ at in the gap, resonant with the edge mode dispersion [see small green dot in Fig.~\ref{fig:band_structure}(a)]. We add this pulse to the r.h.s of Eq.~(\ref{1st_order_eq}),  considering also the dissipation of the condensate with rate $\Gamma$, given by

\begin{align}\label{propagation_eqn}
{i\hbar}\frac{\partial\psi_{x,y,\sigma}}{\partial t} =&\Big(-\Delta e^{-i2\pi\phi_{x,y}\sigma} \psi_{x,y,\bar{\sigma}}\nonumber\\
&+J \psi_{(x-\delta_{1,\bar{\sigma}}),(y-\delta_{1,\sigma}),\sigma}\nonumber\\
&+J \psi_{(x+\delta_{1,\bar{\sigma}}),(y+\delta_{1,\sigma}),\sigma}
-i\Gamma \psi_{x,y,\sigma}+F\Big).
\end{align}

\begin{figure}[htb]
\centering
\includegraphics[width=\linewidth]{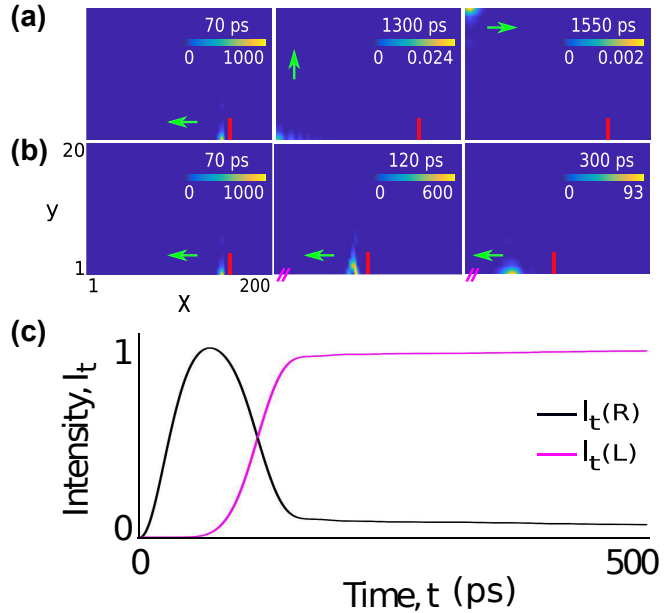}
\caption{(a) Propagation of the one way edge mode. We consider a $200\times20$ lattice formed by micropillars and excite the edge by a coherent probe field focused at the lower edge. We consider $\beta=1/4$, L~$=2$, ${t}_{0}=0$~ps, ${\tau}=60$~ps and choose $k_{p}$ and $\omega_{p}$ from the dispersion curve of the edge states [green dot in Fig.~\ref{fig:band_structure}(a)]. The pulse is injected at $x_{0}=150$ in the lower edge, shown by the red box. The field profiles at different times indicate that the edge state is unidirectional. (b) Propagation of the one-way edge mode in the presence of a defect. We create the defect by removing two micropillars at the lower edge.  Solving the dynamics with the same parameters as above, we plot intensity at different times showing that the edge state goes around the defect and maintains its unidirectionality which emphasizes that the edge state is robust against disorder. (c) Integrated intensity before and after the defect as a function of time, showing that the total intensity is totally transferred from right to left in the long time limit (here the disipationless case is considered to isolate the effects of dispersion and potential scattering from the overall decay).}
\label{fig:one_way}
\end{figure}

Recently much experimental effort was devoted to the implementation of long lifetime  polaritons~\cite{Sun_Nelson,Steger_Snoke}, and we take $\Gamma=250$ ps. Here, we consider the full model as shown in Fig.~\ref{fig:combine_model}(a) of the dimension ($N_{x}\times N_{y}$) and solve the time dynamics to observe the evolution of the states at different times [see Fig.~\ref{fig:one_way}(a)]. In this case the injected pulse is propagating in a single direction along the edges of the rectangular lattice with the absence of backscattering. \par

Now, to demonstrate the robustness of the edge states, we account for the presence of a defect by removing two micropillars along one edge of the rectangular lattice ($N_{x}\times N_{y}$) and exciting the edge mode by injecting the Gaussian pulse as given by Eq.~(\ref{pulse}). From the intensity distribution shown in Fig.~\ref{fig:one_way}(b), we observe that the edge state goes around the defect and maintains its unidirectionality, which demonstrates that the edge state propagation is robust against disorder.\par

To quantify the robustness we calculate the total intensity of the wavepacket, which shall be fully transferred from right to left. We consider the case with no dissipation and plot the total intensity of the sites situated to the right or left of the defect as a function of time [see Fig.~\ref{fig:one_way}(c)]. Here $\text{I}_{\text{t}}(\text{R})$ is the total intensity for the sites on the right side of the defect, and the total intensity of the rest is denoted by $\text{I}_{\text{t}}(\text{L})$. We observe that in the large time limit $\text{I}_{\text{t}}(\text{L})$ slowly reaches the intensity of the pulse before the defect, $\text{I}_{\text{t}}(\text{R})$. Thus, an ultimate limiting factor then corresponds to the lifetime of the pulse.


\subsection*{Topological Insulator lasing mode}
Finally, we make use of the intrinsic driven-dissipative nature of the polaritonic system, and study its lasing properties. It was shown recently that optical analogues of topological insulators can act as powerful single mode lasers, where the influence of disorder is suppressed \cite{Bandres_18_Exp,Harari_Chong_18_theory}. In this section, we show how topological lasing can be achieved in the polaritonic setup, and characterize its gain properties. 

\begin{figure}[htb]
\centering
\includegraphics[width=\linewidth]{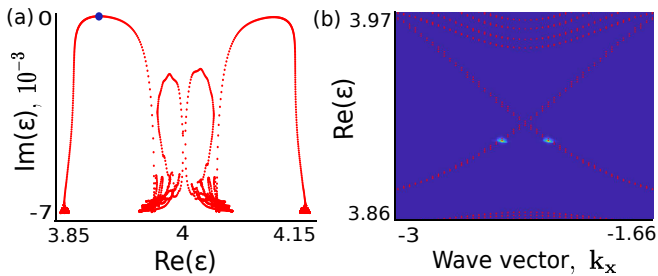}
\caption{ Topological polariton lasing. (a) Imaginary and real part of the modes of the system, calculated from Eq.~\eqref{TI_eqn} for the rectangular lattice ($200\times 20$) formed by micropillars. We consider the system with $\beta=1/4$, $\Gamma=100$~ps, $P=0.0075$~meV. The blue dot points the maximum value of the gain (imaginary part of energy), and corresponds to the edge mode. (b) Zoomed plot of the occupation calculated by evolving Eq.~[\ref{TI_eqn}] in time, averaging over randomly chosen initial conditions. The bright spots correspond to a lasing mode.}
\label{fig:TI_plot}
\end{figure}

We consider the system to be driven with a non-resonant pump, which has a non-zero value along the edges. The time evolution of the system is given by
\begin{align}
\label{TI_eqn}
{i\hbar}\frac{\partial\psi_{x,y,\sigma}}{\partial t} =& \Big(-\Delta e^{-i2\pi\phi_{x,y}\sigma} \psi_{x,y,\bar{\sigma}} \nonumber\\
&+J \psi_{(x-\delta_{1,\bar{\sigma}}),(y-\delta_{1,\sigma}),\sigma}\nonumber\\
&+J \psi_{(x+\delta_{1,\bar{\sigma}}),(y+\delta_{1,\sigma}),\sigma}
-i\Gamma \psi_{x,y,\sigma}\nonumber\\
&+iP\psi_{x,y,\sigma}),
\end{align}
where the last term introduces an incoherent pump of amplitude $P$, as typically done for polaritonic systems in the mean field approximation \cite{CarusottoCiuti}. The gain and loss properties of the system are given by the solution of Eq.~\eqref{TI_eqn} in the frequency domain, which are represented by imaginary and real parts of $\varepsilon$. The results are shown in Fig.~\ref{fig:TI_plot}(a). We find that the eigenvalue with the largest imaginary part, marked by the blue dot in Fig. \ref{fig:TI_plot}(a), corresponds to the edge state. Solving the time dynamics and taking the average over many iterations, we get a large occupation of the edge modes, as shown in Fig.~\ref{fig:TI_plot}(b). Similarly to the previous section, we observe that the edge states are robust against disorder and thus correspond to efficient lasing modes. In analogy to  topological insulator lasers, topological polariton lasing could be realized using electrical injection \cite{Suchomel_Klembt_18} to provide the gain for the lasing mode.

 \section{ Conclusions}

We considered an array of elliptical polariton micropillars in the form of a square lattice. The splitting between the two polarizations of the micropillars, together with interpillar spin dependent tunneling, creates a phase dependent coupling between the neighboring micropillars. With the suitable choice of phases, we are able to generate a uniform flux corresponding to each plaquette, which manifests an artificial uniform magnetic field in the system. This artificial magnetic field represents the breaking of the time reversal symmetry in the system. Considering the toric boundary condition, we calculated the spectrum of the system, which shows fractal behavior akin to the Hofstadter butterfly. Furthermore, using the Dirichlet boundary condition we have observed the edge states, that are unidirectional and robust against disorder. We show that with the non-resonant excitation we can obtain lasing in a topologically protected mode, being a unique feature of the driven-dissipative polaritonic system. This makes the system a promising candidate for practical applications, and puts the proposal on the roadmap for artificial gauge  potential realization with exciton polaritons. 

In this work, we have only considered the linear band structure of exciton-polaritons, neglecting their possible nonlinear interaction. It was recently predicted theoretically that exciton polariton topological insulators using applied ~\cite{Karzig,Bardyn,Nalitov}or effective ~\cite{BardynKarzig2016,Ge2018}, magnetic field could exhibit exhibit bistability ~\cite{Kartashov}, superfluidity~\cite{SigurdssonLi2017}, and solitary waves \cite{Kartashov_2016,Gulevich}. As an outlook we consider a study of nonlinear effects in the presence of artificial gauge fields, and thus open new directions for future research. \par

\section*{Acknowledgments}

We thank the Ministry of education, Singapore (grant no. M0E2017-T2-1-001) for their support. O.K. thanks NTU for hospitality during the initial stage of this work.


\end{document}